\documentclass[12pt]{article}

\usepackage{mathrsfs}
\usepackage[T1]{fontenc}
\usepackage{mathpazo}
\usepackage{setspace}
\usepackage{amsfonts}
\usepackage{amssymb}
\usepackage{amsmath}
\usepackage{epsfig}
\usepackage{latexsym}
\usepackage{color}
\usepackage{graphicx}
\usepackage{nicefrac}
\usepackage[latin1]{inputenc}
\usepackage{pstricks}
\usepackage{slashed}
\usepackage{multirow}

\def\hybrid{\topmargin -20pt    \oddsidemargin 0pt
        \headheight 0pt \headsep 0pt
        \textwidth 6.25in       % A4 paper
        \textheight 9.5in       % A4 paper
        \marginparwidth .875in
        \parskip 5pt plus 1pt   \jot = 1.5ex}

\hybrid

\def\baselinestretch{1.2}

\catcode`\@=11

\def\marginnote#1{}
\newcount\hour
\newcount\minute
\newtoks\amorpm
\hour=\time\divide\hour by60 \minute=\time{\multiply\hour by60
\global\advance\minute by-\hour}
\edef\standardtime{{\ifnum\hour<12 \global\amorpm={am}%
        \else\global\amorpm={pm}\advance\hour by-12 \fi
        \ifnum\hour=0 \hour=12 \fi
        \number\hour:\ifnum\minute<10 0\fi\number\minute\the\amorpm}}
\edef\militarytime{\number\hour:\ifnum\minute<10
0\fi\number\minute}
\def\draftlabel#1{{\@bsphack\if@filesw {\let\thepage\relax
   \xdef\@gtempa{\write\@auxout{\string
      \newlabel{#1}{{\@currentlabel}{\thepage}}}}}\@gtempa
   \if@nobreak \ifvmode\nobreak\fi\fi\fi\@esphack}
        \gdef\@eqnlabel{#1}}
\def\@eqnlabel{}
\def\@vacuum{}
\def\draftmarginnote#1{\marginpar{\raggedright\scriptsize\tt#1}}

\def\draft{\oddsidemargin -.5truein
        \def\@oddfoot{\sl preliminary draft \hfil
        \rm\thepage\hfil\sl\today\quad\militarytime}
        \let\@evenfoot\@oddfoot \overfullrule 3pt
        \let\label=\draftlabel
        \let\marginnote=\draftmarginnote
   \def\@eqnnum{(\theequation)\rlap{\kern\marginparsep\tt\@eqnlabel}%
\global\let\@eqnlabel\@vacuum}  }

\def\preprint{\twocolumn\sloppy\flushbottom\parindent 2em
        \leftmargini 2em\leftmarginv .5em\leftmarginvi .5em
        \oddsidemargin -.5in    \evensidemargin -.5in
        \columnsep .4in \footheight 0pt
        \textwidth 10.in        \topmargin  -.4in
        \headheight 12pt \topskip .4in
        \textheight 6.9in \footskip 0pt
        \def\@oddhead{\thepage\hfil\addtocounter{page}{1}\thepage}
        \let\@evenhead\@oddhead \def\@oddfoot{} \def\@evenfoot{} }

\def\numberbysection{\@addtoreset{equation}{section}
        \def\theequation{\thesection.\arabic{equation}}}

\def\underline#1{\relax\ifmmode\@@underline#1\else
        $\@@underline{\hbox{#1}}$\relax\fi}

\def\titlepage{\@restonecolfalse\if@twocolumn\@restonecoltrue\onecolumn
     \else \newpage \fi \thispagestyle{empty}\c@page\z@
        \def\thefootnote{\fnsymbol{footnote}} }

\def\endtitlepage{\if@restonecol\twocolumn \else \newpage \fi
        \def\thefootnote{\arabic{footnote}}
        \setcounter{footnote}{0}}  %\c@footnote\z@ }

\catcode`@=12 \relax

\def\figcap{\section*{Figure Captions\markboth
        {FIGURECAPTIONS}{FIGURECAPTIONS}}\list
        {Figure \arabic{enumi}:\hfill}{\settowidth\labelwidth{Figure
999:}
        \leftmargin\labelwidth
        \advance\leftmargin\labelsep\usecounter{enumi}}}
 \relax
\def\tablecap{\section*{Table Captions\markboth
        {TABLECAPTIONS}{TABLECAPTIONS}}\list
        {Table \arabic{enumi}:\hfill}{\settowidth\labelwidth{Table
999:}
        \leftmargin\labelwidth
        \advance\leftmargin\labelsep\usecounter{enumi}}}
 \relax
\def\reflist{\section*{References\markboth
        {REFLIST}{REFLIST}}\list
        {[\arabic{enumi}]\hfill}{\settowidth\labelwidth{[999]}
        \leftmargin\labelwidth
        \advance\leftmargin\labelsep\usecounter{enumi}}}
 \relax
\makeatletter
\newcounter{pubctr}
\def\publist{\@ifnextchar[{\@publist}{\@@publist}}
\def\@publist[#1]{\list
        {[\arabic{pubctr}]\hfill}{\settowidth\labelwidth{[999]}
        \leftmargin\labelwidth
        \advance\leftmargin\labelsep
        \@nmbrlisttrue\def\@listctr{pubctr}
        \setcounter{pubctr}{#1}\addtocounter{pubctr}{-1}}}
\def\@@publist{\list
        {[\arabic{pubctr}]\hfill}{\settowidth\labelwidth{[999]}
        \leftmargin\labelwidth
        \advance\leftmargin\labelsep
        \@nmbrlisttrue\def\@listctr{pubctr}}}
 \relax
\makeatother
\newskip\humongous \humongous=0pt plus 1000pt minus 1000pt

\newif\ifdtup

\relax

\def\be{\begin{equation}}
\def\ee{\end{equation}}
\def\ba{\begin{eqnarray}}
\def\ea{\end{eqnarray}}

\begin{document}
%\draft

%\renewcommand{\theequation}{\arabic{equation}}
\renewcommand{\theequation}{\thesection.\arabic{equation}}

\newcommand{\beq}{\begin{equation}}
\newcommand{\eeq}[1]{\label{#1}\end{equation}}
\newcommand{\ber}{\begin{eqnarray}}
\newcommand{\eer}[1]{\label{#1}\end{eqnarray}}
\newcommand{\eqn}[1]{(\ref{#1})}
\begin{titlepage}
\begin{center}

%\hfill hep--th/yymmnnn\\
\vskip -.1 cm
\hfill October 2011\\

\vskip .4in

{\Large \bf More on axial anomalies of Lifshitz fermions}
%{\Large \bf The $\eta$-invariant of certain higher order operators on $S^3$}

\vskip 0.5in

{\bf Ioannis Bakas} \vskip 0.2in
{\em Department of Physics, University of Patras \\
GR-26500 Patras, Greece\\
\vskip 0.2in
\footnotesize{\tt bakas@ajax.physics.upatras.gr}}\\

\end{center}

\vskip 0.6in

\centerline{\bf Abstract}
\vskip 0.2in
\noindent
We show that the gauge and metric field contribution to the axial anomaly of
a four-dimensional massless Lifshitz fermion theory with anisotropy scaling exponent
$z$ is identical to the relativistic case, hereby extending the results found in
arXiv:1103.5693 to arbitrary values of $z$. This is in accordance with the fact
that the axial anomaly is an infra-red phenomenon in disguise. We also provide
some new models that realize baryon and lepton number violation in non-relativistic
theories of gravity. In all cases, the volume of space exhibits a lower bound
that is fixed by the gravitational coupling parameters.
\vfill
\end{titlepage}
\eject

\def\baselinestretch{1.2}
\baselineskip 16 pt \noindent

\section{Introduction}
\setcounter{equation}{0}

It has been known for a long time that a massless Dirac fermion in four space-time dimensions
exhibits a quantum anomaly in the axial current conservation law when coupled to external
gauge and/or metric fields. The anomalous term in the divergence of the axial current
$J_5^{\mu} = \bar{\Psi} \gamma^{\mu} \gamma_5 \Psi$ is proportional to the characteristic classes
${\rm Tr}(F \wedge F)$ and ${\rm Tr}(R \wedge R)$ written in terms of the curvature two-forms
$F$ and ${R^a}_b$ of the gauge and metric fields, respectively, with corresponding coefficients
$1/4\pi^2$ and $1/96 \pi^2$ (in appropriate units that will be used throughout this paper)
\cite{bell, salam}. The coefficient of anomaly is half of that when computed for a Weyl
fermion. This result describes the local form of the axial anomaly, whereas its integral
over space-time has been used to study possible violations of chiral symmetry via
the Atiyah-Singer index theorem (for an overview of the subject from a physicist's perspective
see, for instance, \cite{eguchi} and references therein). The axial charge $Q_5$ is not
conserved when the four-dimensional Dirac operator in a given background exhibits an asymmetry
in the number of positive and negative chirality zero modes, leading to violations of
lepton and baryon number in the theory \cite{hooft}. Generalizations to higher dimensions have
also been worked out systematically \cite{witten} and the results found many important
applications in physics.

In this paper we examine the occurrence of axial anomalies in the non-relativistic fermion
field theories of Lifshitz type. For this we consider higher order generalizations
of the Dirac operator of the form $Q = i \gamma^0 \partial_0 + i \gamma^i \partial_i
(- \partial_k \partial^k + M^2)^{\alpha}$ acting on spinors $\Psi$, where $M$ is an arbitrary
mass scale and $\alpha$ is a free parameter. The resulting theory of Lifshitz fermions, which
is described by the Lagrangian density $\bar{\Psi} Q \Psi$ exhibits anisotropic scaling in
space and time by assigning dimensions $[L] = -1$, $[T] = -z$ and $[\Psi] = z/2$ with
parameter $z= 2\alpha + 1$. This model has an axial current which is conserved at the classical
level. Its time component $J_5^0 = \bar{\Psi} \gamma^0 \gamma_5 \Psi$ is the same as in the
relativistic case (the same also applies to the axial charge $Q_5$), but the spatial components
$J_5^i$ are more complicated as they involve derivative terms that depend upon $\alpha$; their
precise form will not be needed for the
purposes of the present work. Then, a natural question is whether there are anomalies in the
axial current conservation law that arise upon quantization of the fermions in the background
of gauge and/or metric fields, and, if so, what will be their form compared to the relativistic
case, $\alpha = 0$. The result turns out to be independent of the parameters $M$ and $\alpha$,
and, in this sense, the structure of the axial anomalies appears to be universal.

The present work generalizes earlier results on the subject that were obtained for Lifshitz
fermions with anisotropy scaling parameter $z=3$ coupled to background gauge fields
\cite{wadia} and gravity \cite{dieter} in four space-time dimensions. Here, $z = 2 \alpha + 1$
is left arbitrary and it can even assume fractional values. For practical reasons we restrict
attention to four space-time dimensions, although we expect the same universality to govern the
form of the axial anomalies in higher dimensional Lifshitz fermion models as in relativistic
theories \cite{witten}. An intuitive explanation for all this is provided by the fact that the
axial anomaly is an infra-red phenomenon in disguise, and, hence, any higher order corrections
to the Dirac operator that become relevant in the ultra-violet regime ought to leave its form
unaltered (the way it works in relativistic theories is nicely explained in \cite{witten},
but see also \cite{hooft2, adam, cole} for earlier work on the subject).
The introduction of the mass scale $M$ into the definition of $Q$
serves precisely the purpose of showing by explicit computation that the axial anomaly is
indeed independent of it and the order of the higher derivative corrections to the fermion
operator. The result should be contrasted to the form of the Weyl anomaly in Lifshitz
theories that exhibit anisotropic local scale invariance. In that case one does not expect
the result to be the same as for relativistic fields because the Weyl anomaly is an
ultra-violet phenomenon that is sensitive to higher order corrections at short distances.
Some partial results in this direction can be found in \cite{theisen} for scalar field models,
whereas generalizations to other field theories of Lifshitz type have not been addressed
so far in the literature.

Thus, the computation of axial anomalies in non-relativistic massless
fermion models is a much simpler problem compared to other type of anomalies and at the same
time it has many interesting applications related to chiral symmetry breaking at a Lifshitz
point. A closely related problem is the parity anomaly of three-dimensional fermion theories
coupled to external gauge and/or metric fields. The methods we use in this paper suffice to
demonstrate the universality of the parity violating piece of the effective action induced by
massless fermions regardless the order of the three-dimensional fermion operator. This
result is a manifestation of the close relation that exists between different type of
anomalies in odd and even number of space-time dimensions generalizing the framework that
was established long time ago for relativistic theories (see, for instance, \cite{moore}
and references therein). We will say more about this later focusing, in particular, to
the gravitational case and the universal character of the induced three-dimensional
Chern-Simons action.

In section 2, we employ the path integral method to extract the local form of the axial
anomaly of Lifshitz fermion models. The computation is performed rather easily for gauge
field backgrounds and the result is identical to the relativistic case, as expected on general
grounds. Confirming the universality of the axial anomaly for metric field backgrounds is much
more involved computationally by path integral methods. For this reason we employ an indirect
method based on the integrated form of the anomaly and the associated index theorem for the
Dirac-Lifshitz operator. Then, using this framework, the coefficient of the axial anomaly follows
from the computation of the $\eta$-invariant of the associated three-dimensional Lifshitz operator
on certain geometries. In section 3, we briefly review the notion of $\eta$-invariant and its
relation to anomalies in relativistic fermion theories in three and four dimensions.
In section 4, we compute the $\eta$-invariant of higher order fermion operators on homogeneous
geometries and show that the result is independent of the parameters $M$ and $\alpha$. This
suffices to prove our claim about the gravitational contribution to the axial anomaly of
Lifshitz fermion. In section 5, we present some applications to gravitational theories of
Lifshitz type and examine the possibility of chiral symmetry breaking by instantons, thus
generalizing the results found in \cite{dieter}. Finally, section 6 contains our conclusions.

\section{Path integral derivation of axial anomaly}
\setcounter{equation}{0}

Coupling the Lifshitz fermion theory to external fields amounts to replacing
$\partial_{\mu}$ by $D_{\mu}$, as usual. Then, the interacting Dirac-Lifshitz operator
takes the form
\be
{\cal Q} = i \gamma^{\mu} {\cal D}_{\mu} = i \gamma^0 D_0 + {1 \over 2} i \gamma^i
[D_i (- D_k D^k + M^2)^{\alpha} + (- D_k D^k + M^2)^{\alpha} D_i]
\label{aroura}
\ee
up to a factor ordering ambiguity that turns out to be irrelevant for the computation of
the anomaly. To avoid unnecessary complications or ambiguities in the definition of
the operator \eqn{aroura}, and in view of the applications that will be discussed later,
coupling to geometry is taken with respect to metrics $-(dx^0)^2 + g_{ij}(x^0, x) dx^i dx^j$.
Such backgrounds arise naturally in the canonical decomposition of the metric field in
space-times of the form $I \times \Sigma_3$ choosing the lapse and shift functions as
$N=1$ and $N_i = 0$.

The anomalous divergence of the axial current can be found by applying Noether's
procedure to the fermionic path integral of the interacting theory, which is most
conveniently described in the Euclidean domain after Wick rotation of time
$x^0 \rightarrow it$. The measure $({\cal D} \bar{\Psi}) ({\cal D} \Psi)$ is not invariant
under chiral rotations $\delta_{\epsilon} \Psi = i \epsilon \gamma_5 \Psi$, giving rise
to a non-trivial contribution to the divergence of the axial current (see, for instance,
\cite{suzuki} and references therein to the original papers). Using the eigen-states
$\varphi_n (t, x)$ of the interacting fermion operator $i \gamma^{\mu} {\cal D}_{\mu}$,
the result takes the intermediate form
\be
\nabla_{\mu} J_5^{\mu} (t, x) = 2 \lim_{\Lambda \rightarrow \infty} \sum_n \varphi_n^{\dagger}
(t, x) \gamma_5 ~ e^{- (i\gamma^{\mu} {\cal D}_{\mu})^2 / \Lambda^{2z}} \varphi_n (t, x)
\label{lirak}
\ee
after introducing a cut-off $\Lambda$ to regulate the infinite sum that otherwise is
ill-defined. Note that $\Lambda$ is raised to the power $2z$ to match the scaling properties
of the square of the Dirac-Lifshitz operator for general values of $z$.

The gauge field contribution to the axial anomaly can be computed relatively easy for all
values of the anisotropy scaling parameter $z$. First, using the algebra of the Dirac
gamma-matrices, it is convenient to rewrite the square of the Dirac-Lifshitz operator
appearing in the regulator as
\be
(i\gamma^{\mu} {\cal D}_{\mu})^2 = - {\cal D}_{\mu} {\cal D}^{\mu} - {1 \over 4}
[\gamma^{\mu} , ~ \gamma^{\nu}] [{\cal D}_{\mu} , ~ {\cal D}_{\nu}] ~,
\label{decompo}
\ee
where $D_{\mu} = \partial_{\mu} - i A_{\mu}$ provides the minimal coupling of the theory
to an external gauge field $A_{\mu}$, which can be Abelian or non-Abelian. Next, using
the plane wave basis of solutions of the free Dirac-Lifshitz operator,
${\rm exp}(ik_{\mu}x^{\mu})$, we can extract the explicit gauge field dependence of the
anomalous term that appears on the right-hand side of equation \eqn{lirak}. Since
the action of the interacting Dirac-Lifshitz operator on plane waves amounts to replacing
$D_{\mu}$ by $D_{\mu} + i k_{\mu}$ everywhere, it follows that the anomalous divergence of
the axial current in a gauge field background takes the form,
\ba
\nabla_{\mu} J_5^{\mu} & = & 2 \lim_{\Lambda \rightarrow \infty} \Lambda^{z+3} ~
{\rm Tr} \int {d^4 k \over (2\pi)^4} ~ e^{-k_0^2 - k^{2z}} ~ \gamma_5 ~ {\rm exp}
\Big\{-{i ~ k^{2 \alpha} \over 2 \Lambda^{z+1}} [\gamma^0 , ~ \gamma^i] \left(F_{0i} +
2\alpha F_{0l} \hat{k}_i \hat{k}^l \right)  \nonumber\\
& &  ~~~~~~~~~ - {i ~ k^{4\alpha} \over 4 \Lambda^2} [\gamma^j , ~ \gamma^k] \left(F_{jk} +
4\alpha F_{jl} \hat{k}_k \hat{k}^l \right) + \cdots \Big\} ~ .
\ea

In writing this equation we have also rescaled $k_0$ to $\Lambda^z k_0$ and $k_i$ to
$\Lambda k_i$ conforming with the anisotropic scaling of the components of the momentum
vector $k_{\mu} = (k_0, ~ k_i)$ that is implied by the scaling of the space-time coordinates
$x_{\mu} = (x_0 , ~ x_i)$ and introduced the unit three-momentum vector with components
$\hat{k}_i$, setting $k_i = k \hat{k}_i$, where $k^2 = k_i k^i$. Here, we have depicted the
most relevant terms that originate from the exponential of
$(i \gamma^{\mu} {\cal D}_{\mu})^2$ after replacing $D_{\mu}$ by $D_{\mu} + i k_{\mu}$.
The factor ${\rm exp}(-k_0^2 - k^{2z})$ originates from ${\cal D}_{\mu} {\cal D}^{\mu}$ in
the decomposition \eqn{decompo}, which is taken in the Euclidean domain, whereas the two
terms shown explicitly in curly brackets originate from the electric and magnetic components
$[\gamma^0 , ~ \gamma^i][{\cal D}_0 , ~ {\cal D}_i]$ and
$[\gamma^j , ~ \gamma^k][{\cal D}_j , ~ {\cal D}_k]$, respectively, appearing on the
right-hand side of \eqn{decompo}. The terms that are omitted carry explicit dependence on
the mass scale $M$, but they do not contribute to the final result. Such terms yield
contributions that either vanish as $\Lambda \rightarrow \infty$ or their trace is zero
because they do not contain sufficient number of gamma-matrices multiplied with $\gamma_5$.

The calculation proceed by expanding the exponential in curly brackets and noting that
the only relevant terms arise to second order. We also perform the Gaussian integration
over $k_0$, picking up a factor of $\sqrt{\pi}$, and use the identity
${\rm Tr}(\gamma_5 [\gamma^0 , ~ \gamma^i] [\gamma^j , ~ \gamma^k]) = - 16 \epsilon^{0ijk}$.
Then, after taking the limit $\Lambda \rightarrow \infty$, we obtain
\be
\nabla_{\mu} J_5^{\mu} = {1 \over 4 \pi^{7/2}} \epsilon^{0ijk} \int d^3 k e^{-k^{2z}}
k^{3z - 3} {\rm Tr} \left(F_{0i} F_{jk} + 2 \alpha F_{0l} F_{jk} \hat{k}_i \hat{k}^l
+ 4 \alpha F_{0i} F_{jl} \hat{k}_k \hat{k}^l \right) ,
\ee
setting $z=2 \alpha + 1$. The calculation is completed by integrating out the three-momenta.
Using polar coordinates in momentum space with $\hat{k} = ({\rm sin} \theta {\rm sin} \phi ,
~ {\rm sin} \theta {\rm cos} \phi , ~ {\rm cos} \theta)$ and substituting $x = k^z$, it follows
easily that
\be
\int d^3 k ~ e^{-k^{2z}} k^{3z-3} = 4\pi \int_0^{\infty} dk ~ k^{3z-1} e^{-k^{2z}} =
{1 \over z} \pi^{3/2}
\ee
and
\be
\int d^3 k ~ e^{-k^{2z}} k^{3z-3} ~ \hat{k}_i \hat{k}_j = {4 \pi \over 3}
~ \delta_{ij} \int_0^{\infty} dk ~ k^{3z-1} e^{-k^{2z}} =
{1 \over 3z} \pi^{3/2} \delta_{ij} ~.
\ee
The final result is independent of the scaling parameter $z$,
\be
\nabla_{\mu} J_5^{\mu} = {1 \over 4 \pi^2} \epsilon^{0ijk} {\rm Tr} (F_{0i} F_{jk}) =
{1 \over 4 \pi^2} {\rm Tr} (F \wedge F) ~,
\ee
and coincides with the anomaly of the axial current conservation law of a Dirac fermion.

Similarly, we may compute the gravitational contribution to the axial anomaly under
the minimal substitution of $\partial_{\mu}$ by
$D_{\mu} = \partial_{\mu} + (1/8) [\gamma_a , ~ \gamma_b] {\omega_{\mu}}^{ab}$ written
in terms of the spin connection $\omega$.  The regulated sum shown in \eqn{lirak} is
much more difficult to evaluate in this case as one has to introduce point splitting and
use the curved space analogue of plane waves given in terms of the geodesic interval in
space-time. Keeping track of the relevant terms is also quite tricky, in general, apart from
the $z=1$ case that corresponds to Dirac fermions. For higher values of $z$ one encounters
multiple derivatives of the geodesic interval which in the coincidence limit correspond
to Synge-DeWitt tensors of rank bigger than $2$. There are many sources of such terms
that contribute to the final result and one has to sum them up carefully. For $z=3$ the
intermediate calculations are already quite tedious (see \cite{dieter} for details)
and one expects that this method will be practically impossible to carry out to the end
for general values of $z$. A closely related (but practically simpler) calculational
method could be developed by applying path integral methods to suitably chosen models
of supersymmetric quantum mechanics with Hamiltonian equal to the square of the
Dirac-Lifshitz fermion operator. This method was successfully employed in the past to
compute the form of anomalies in relativistic theories \cite{witten}, particularly in
higher dimensions where the conventional approach becomes intractable, but it is not
currently known how to generalize it to encompass non-relativistic fermion systems. Thus,
we should resort to other means for the efficient calculation of the gravitational
contribution to the axial anomaly for general values of the anisotropy scaling parameter
$z$.

The axial anomaly is necessarily a total derivative term, as it obstructs the divergence
of a current. It is also gauge invariant, meaning that the anomalous term is solely expressed
via the field strength $F$ and/or $R$. In four dimensions there is only one such term given by
the topological density ${\rm Tr} (F \wedge F)$ when fermions couple to gauge fields.
Thus, the whole purpose of the path integral computation is to determine the coefficient
of this term, which, as we saw before, turns out to be universal. By the same token,
when fermions couple to metric fields, there are two possible terms given by the topological
densities ${\rm Tr} (R \wedge R) = {R^a}_b \wedge {R^b}_a$ and
${\rm Tr} (R \wedge {}^* R) = (-1/2)\epsilon_{abcd} {R^a}_b \wedge {R^c}_d$ in four dimensions.
The second term is easily ruled out by the trace identities of Dirac gamma-matrices;
close inspection of the anomalous term \eqn{lirak} reveals that the trace cannot yield
more that one epsilon symbol, and, hence only ${\rm Tr} (R \wedge R)$ and not
${\rm Tr} (R \wedge {}^* R)$ can possibly contribute to the axial anomaly. Then, the
only remaining task is to evaluate its coefficient and verify that in all cases the
result is the same, namely
\be
\nabla_{\mu} J_5^{\mu} = - {1 \over 96 \pi^2} \epsilon^{0ijk}
{R^{ab}}_{0i} R_{ab ~ jk}
= {1 \over 96 \pi^2} {\rm Tr} (R \wedge R) ~.
\label{tsoli}
\ee

The method that we will follow relies on the integrated form of the anomaly and
its relation to the index theorem of the Dirac-Lifshitz operator. An important
ingredient in the computation is the $\eta$-invariant of the fermion operator
restricted on the three-dimensional leaves of space-time foliation and its intimate
relation to the gravitational Chern-Simons action via the anomaly, which turns out
to be universal.

\section{The $\eta$-invariant and axial anomalies}
\setcounter{equation}{0}

In this section we make a small detour to summarize some basic facts about the index theorem
of the Dirac operator and the associated notion of $\eta$-invariant, which play important role
in relativistic quantum field theory. Later we will extend the results to the more general
class of Dirac-Lifshitz fermion operators and use them to determine the coefficient of the
axial anomaly by spectral methods.

Recall that the $\eta$-invariant of a self-adjoint matrix $A$ with real eigen-values
$\lambda_k$ is the signature of $A$, which is defined to be
\be
\eta_{A} = {\rm sign} (A) =  {\sum_{\lambda_k}}^{~ \prime} ({\rm sign} ~ \lambda_k) =
\sum_{\lambda_k > 0} 1 - \sum_{\lambda_k < 0} 1 ~,
\ee
assuming that there are no zero eigen-values in the spectrum, which otherwise are excluded.
As such, $\eta_{A}$ is invariant under rescaling of the matrix by any positive number
and it counts the spectral asymmetry between positive and negative
eigen-values of $A$. If $A$ depends upon a parameter $\delta$, the spectrum will flow as
$\delta$ varies continuously and level crossing may occur in the system. The $\eta$-invariant
of $A (\delta)$ changes by $\pm 2$ units every time a negative (respectively positive)
eigen-value at $\delta_0$ crosses to positive (respectively negative) values at $\delta_1$
as $\delta$ varies from $\delta_0$ to $\delta_1$.

Computing the $\eta$-invariant is a notoriously difficult task when the matrix $A$ is
infinite dimensional, since the spectrum of $A$ cannot be determined in general and the
definition of $\eta_A$ becomes ill-defined as it stands. The standard prescription in those
cases is to consider first the $\eta$-function of $A$ obtained by zeta-function regularization,
as
\be
\eta_A (s) =  {\sum_{\lambda_k}}^{~ \prime} ({\rm sign} ~ \lambda_k) |\lambda_k|^{-s} ,
\ee
and then define the corresponding $\eta$-invariant as
\be
\eta_A = \eta_A (0) ~.
\ee
Typically, $A$ is a self-adjoint operator on a compact manifold so that its spectrum is
discrete and $\eta_A (s)$ can be analytically extended to a meromorphic function on the
entire complex $s$-plane which is regular at $s=0$.
Yet, direct computation of the $\eta$-invariant by spectral methods can only be done in
a few cases. Such examples will be given later.

The Atiyah-Patodi-Singer index theorem is the main mathematical framework that necessitated
the introduction of the $\eta$-invariant of various operators \cite{APS} (but see also
the textbooks \cite{gilka, melrose} for more details). Restricting attention to the Dirac
operator $i \gamma^{\mu} D_{\mu}$ on a four-dimensional Riemannian manifold $M_4$, the
difference of its positive and negative chirality zero modes, known as the index of
$i \gamma^{\mu} D_{\mu}$ (we will use $D$ in short), is given by
\be
{\rm Ind}(D) = - {1 \over 192 \pi^2} \int_{M_4} {\rm Tr} (R \wedge R) +
{1 \over 192 \pi^2} \int_{\partial M_{4}} {\rm Tr} (\theta \wedge R) - {1 \over 2}
\eta_{\rm D} (\partial M_4) ~.
\label{aps}
\ee
The first term provides the bulk contribution to the index, which is obtained by integration
over $M_4$ of (a half times) the local form of the axial anomaly of a Dirac fermion in the
background of a metric field.
It involves the (appropriately normalized) characteristic class ${\rm Tr} (R \wedge R)$
that is written in terms of the curvature two-form $R = d \omega + \omega \wedge \omega$
associated to the spin-connection $\omega$ of the metric $g$ on $M_4$. The other two terms
arise when $M_4$ has a boundary $\partial M_4$. The local boundary term is the integral over
$\partial M_4$ of the (appropriately normalized) secondary characteristic class
${\rm Tr} (\theta \wedge R)$, \cite{simons}, which is written in terms of the second fundamental
form $\theta = \omega - \omega^0$ and accounts for the possible deviation of the metric
$g$ from a cross-product form $g_0$ at the boundary. Finally, the last term is a non-local
boundary contribution to the index provided by the $\eta$-invariant of the tangential part
of the Dirac operator restricted to $\partial M_4$. Here, it is implicitly assumed that
the four-component spinors $\Psi$ have a specific fall-off rate close to $\partial M_4$
(known as APS boundary conditions) and so what we are really computing is the $L^2$-index of
normalizable zero modes of $i \gamma^{\mu} D_{\mu}$.

An immediate consequence of the Atiyah-Patodi-Singer index theorem is that the $\eta$-invariant
of the three-dimensional Dirac operator on a compact manifold $\Sigma_3$ without boundary,
which is endowed with a Riemannian metric, is related to the gravitational Chern-Simons action
\cite{simons}
\be
W_{\rm CS} (\Sigma_3) = \int_{\Sigma_3} {\rm Tr} \left(\omega \wedge d \omega + {2 \over 3}
\omega \wedge \omega \wedge \omega \right)
\label{csact}
\ee
as follows,
\be
{1 \over 2} \eta_{\rm D} (\Sigma_3) + {1 \over 192 \pi^2} W_{\rm CS} (\Sigma_3) =
{\rm constant} ~.
\label{moulia}
\ee
This is easily seen by considering a four-dimensional spin manifold $M_4$ bounded by
$\Sigma_3$ and choosing a metric on it that has cross-product form near the boundary
$\partial M_4 = \Sigma_3$. Thus, although both the $\eta$-invariant and the Chern-Simons
action depend explicitly on the conformal class of the metric on $\Sigma_3$, their
(appropriately weighted) sum is in fact independent of it. The coefficient of $W_{\rm CS}$
is uniquely fixed by the anomaly. We also note for completeness
that both $\eta_{\rm D}$ and $W_{\rm CS}$ are odd under parity, thus flipping sign under
orientation reversing diffeomorphisms on the manifold $\Sigma_3$.

Equation \eqn{moulia} provides an indirect way to compute the $\eta$ invariant
of the three-dimensional Dirac operator, up to a constant, based on axial anomalies in four
space-time dimensions. Conversely, one may use this formula to extract the coefficient
of the anomalous term ${\rm Tr} (R \wedge R)$ in the divergence of the axial current using
spectral methods, provided that the $\eta$-invariant can be computed explicitly for a
certain class of three-metrics. This is the strategy that will also be applied later
to Lifshitz models.

Alternatively, on a four-dimensional manifold of the form $M_4 \simeq I \times \Sigma_3$,
where $I$ is an interval (it can also be $I \simeq \mathbb{R}$ in the physical applications)
endowed with the metric
\be
ds^2 = dt^2 + g_{ij}(t, x) dx^i dx^j ~,
\label{spatim}
\ee
the Atiyah-Patodi-Singer index theorem for the Dirac operator implies the following
useful relation
\be
{1 \over 2} \Delta \eta_{\rm D} (\Sigma_3) + {1 \over 192 \pi^2} \Delta W_{\rm CS} (\Sigma_3)
= {\rm integer} ~,
\label{rikoti}
\ee
where $\Delta$ denotes the difference of any given quantity at the two end-points of the
Euclidean time interval $I$. One may view \eqn{moulia} as special case of the more general
relation \eqn{rikoti} by just shrinking one of the two ends of the cylinder $I \times \Sigma_3$
to a point. For example, for $\Sigma_3 \simeq S^3$, let us assume without loss of generality
that one of the two boundaries is endowed with the constant curvature (round) metric. At that
boundary the $\eta$-invariant vanishes, as there can be no asymmetry in the spectrum of the
Dirac operator, whereas $W_{\rm CS}$ assumes the value $16 \pi^2$. Shrinking that boundary
to a point by conformal rescaling does not affect the values of $\eta_{\rm D}$ and
$W_{\rm CS}$ and \eqn{rikoti} reduces to \eqn{moulia} with constant term equal to
$16 \pi^2 / 192 \pi^2 = 1/12$ modulo integers.

The integer appearing on the right-hand side of equation \eqn{rikoti} is the
index of the Dirac operator on $I \times \Sigma_3$. If the space-time metric \eqn{spatim} has
a cross-product form everywhere, i.e., $g_{ij}$ is independent of $t$, the three-dimensional
metric at the two ends of the cylinder (boundaries of space-time) will be the same, and, hence,
$\Delta \eta_{\rm D} = 0 = \Delta W_{\rm CS}$ giving zero index. Otherwise, if the metric
is warped by depending explicitly on $t$, the $\eta$-invariant and the Chern-Simons action
will be different at the two end-points in general. Furthermore, it can be shown in this case
that the index of the four-dimensional Dirac operator is provided by the spectral flow,
\be
{\rm Ind}(D) = \Delta S (\Sigma_3) ~,
\ee
where $\Delta S (\Sigma_3)$ is the net number of level crossings that may occur in the
spectrum of the three-dimensional Dirac operator on $\Sigma_3$ as the metric $g_{ij} (t, x)$
deforms from one end-point of the interval $I$ to the other end-point (recall that the
$\eta$-invariant jumps by $\pm 2$ units when an eigen-value crosses from negative to
positive values or conversely, and, therefore, each level crossing contributes $\pm 1$ units
to the index). This formula and its generalization to Lifshitz models is very useful
for the applications that will be discussed later.

There is another occurrence of the $\eta$-invariant of the three-dimensional Dirac operator
in quantum field theory. This time, one is interested in extracting the parity-violating piece
of the effective action $\Gamma_{\rm eff}$,
\be
{\rm exp} \left(- \Gamma_{\rm eff} \right) = \int ({\cal D} \bar{\Psi}) ({\cal D} \Psi)
\left(- \int d^3x ~ \bar{\Psi} (x) i \gamma^i D_i \Psi (x) \right) ,
\ee
which is obtained by integrating out the fermions. Here, the Dirac operator is minimally
coupled to external fields, such as gauge and/or metric fields, and, therefore, the
fermion effective action depends explicitly on them. There is a natural relation between
the parity anomaly in odd dimensions and the gauge and/or gravitational contribution to
axial anomalies in even dimensions \cite{moore}. In our case, the imaginary part
of the effective action that provides the parity violating piece is
\be
{\rm Im} \Gamma_{\rm eff} = {\pi \over 2} \eta_{\rm D} ~,
\label{mitia}
\ee
which in turn is related to the parity violating Chern-Simons action. The derivation is
based on the formal relation (choosing the branch ${\rm log}(-1) = i \pi$)
\be
{1 \over 2} {\rm log} {{\rm det} (- A) \over {\rm det} (A)} = {i \pi \over 2} \eta_A
\ee
and it can be made rigorous by zeta-function regularization of the determinant of the
Dirac operator \cite{reuter}. Thus, the Chern-Simons term of topologically massive gauge
and/or gravitational theories in three dimensions is induced radiatively
\cite{redlich, niemi, vuorio} (but see also \cite{jackiw} for earlier work on the subject).
A closely related subject is the structure of the induced anomalous vacuum current in odd
dimensions and the $\eta$-function regularization of the vacuum charge of a fermion field
(see, for instance, \cite{lott} for a general and rigorous discussion of this subject).

\section{Universality of $\eta$-invariant in Lifshitz models}
\setcounter{equation}{0}

The index of the Dirac-Lifshitz operator $i \gamma^{\mu} {\cal D}_{\mu}$ defined by
\eqn{aroura} (we will use ${\cal D}$ in short) follows from the integrated form of
the axial anomaly. Thus, on a space-time $M_4$ with boundaries we have
\be
{\rm Ind}({\cal D}) = - {1 \over 192 \pi^2} \int_{M_4} {\rm Tr} (R \wedge R) +
{1 \over 192 \pi^2} \int_{\partial M_{4}} {\rm Tr} (\theta \wedge R) - {1 \over 2}
\eta_{\mathbb{D}} (\partial M_4) ~,
\label{aps2}
\ee
in exact analogy with the index of the Dirac operator \eqn{aps}, provided that the
coefficient of the anomalous term in \eqn{tsoli} is the same as for the Dirac operator
in a metric background. Here, the $\eta$-invariant refers to the operator of order $z$
\be
i \gamma^i \mathbb{D}_i = {1 \over 2} i \gamma^i
[D_i (- D_k D^k + M^2)^{\alpha} + (- D_k D^k + M^2)^{\alpha} D_i]
\ee
acting on two-component spinors on the three-dimensional boundary $\partial M_4$.

As before, the Atiyah-Patodi-Singer theorem \eqn{aps2} implies the following
identity between $\eta_{\mathbb{D}}$ and the gravitational
Chern-Simons action on a compact three-manifold $\Sigma_3$ without boundaries
\be
{1 \over 2} \eta_{\mathbb{D}} (\Sigma_3) + {1 \over 192 \pi^2} W_{\rm CS} (\Sigma_3) =
{\rm constant} ~.
\label{moulia2}
\ee
Then, comparison with equation \eqn{moulia} for the Dirac operator implies that
$\eta_{\mathbb{D}} (\Sigma_3)$ ought to be equal to $\eta_{\rm D} (\Sigma_3)$ up to
a constant. For $\Sigma_3 \simeq S^3$, in particular, this constant has to be zero
because $S^3$ admits a constant curvature metric for which there can be no spectral
asymmetry for $D$ nor $\mathbb{D}$. Conversely, proving that the $\eta$-invariant
is universal, i.e.,
\be
\eta_{\mathbb{D}} (S^3) = \eta_{\rm D} (S^3)
\label{eetaa}
\ee
for any given metric on $S^3$, suffices to show that the coefficient of the
anomaly in the axial current conservation law \eqn{tsoli} is indeed $-1/192 \pi^2$.
This is precisely what we will show next by restricting attention to a particular
class of metrics that depend continuously on a particular modulus (anisotropy parameter).
Hence, we will provide an alternative derivation of the anomaly by spectral methods, as
advertised before.

The class of metrics that will be used in the following allow for exact computation
of the $\eta$-invariant. Then, as consequence of the Atiyah-Patodi-Singer theorem,
equation \eqn{eetaa} should hold true for all metrics on $S^3$. This way we will also be
able to provide model geometries for which the index of the four-dimensional fermion
operator can be computed explicitly by spectral flow. Finally, we will derive geometric
conditions for having non-zero index on certain backgrounds. The presentation we
follow here has many common elements to our earlier work \cite{dieter}, but it is
now applicable to Lifshitz operators of arbitrary order $z$.

First, to set up the stage, we consider some general aspects of the eigen--value problem
of the Dirac operator $i \gamma^i D_i$ on a three-dimensional manifold $\Sigma_3$ with
a Riemannian metric $g$ acting on two--component spinors,
\be
i \gamma^i D_i \Psi (x) = \zeta ~ \Psi (x) ~ ; ~~~~~~
\Psi (x) = \left(\begin{array}{c}
\psi_1 \\
 \\
\psi_2 \\
\end{array} \right) ,
\ee
where
\be
i \gamma^i D_i = i \gamma^I {E_I}^i \left(\partial_i + {1 \over 8}
[\gamma_J , ~ \gamma_K] {\omega^{JK}}_i \right) .
\ee
Here, ${E_I}^i$ are the components of the inverse dreibeins associated to $g_{ij}$ with
tangent space indices $I$ and ${\omega^{JK}}_i$ are the components of the corresponding spin
connection. Also, the gamma matrices $\gamma^I$ are chosen in terms of the Pauli matrices,
as usual,
\be
\gamma^1 = \left(\begin{array}{ccc}
0 &  & 1 \\
  &  &   \\
1 &  & 0
\end{array} \right) , ~~~~~
\gamma^2 = \left(\begin{array}{ccc}
0 &  & i \\
  &  &   \\
-i &  & 0
\end{array} \right) , ~~~~~
\gamma^3 = \left(\begin{array}{ccc}
1 &  & 0 \\
  &  &   \\
0 &  & -1
\end{array} \right)
\ee
and satisfy the anti-commutation relations $[\gamma^I , \gamma^J]_+ = 2 \delta^{IJ}$.
A useful relationship is provided by Lichnerowicz's formula for the square of the
Dirac operator \cite{lichn},
\be
(i \gamma^i D_i)^2 = - D^2 + {1 \over 4} R ~,
\label{lichn}
\ee
which is written in terms of the Bochner Laplacian operator $D^2 = D_i D^i$ acting on
spinors (rather than scalars) and the Ricci scalar curvature $R$ associated to the
three-dimensional metric $g$.

The eigen-value problem for the more general class of operators $i \gamma^i \mathbb{D}_i$ of
order $z= 2 \alpha + 1$,
\be
i \gamma^i \mathbb{D}_i \Psi (x) = Z ~ \Psi (x) ~,
\ee
is definitely a more complex problem. Even if the spectrum $\zeta$ of the Dirac operator can
be determined for a given metric $g$, the eigen-values $Z$ will not be easy to obtain in general.
Note at this point, based on Lichnerowicz's formula \eqn{lichn}, that we may express the operators
$i \gamma^i \mathbb{D}_i$ in the form
\be
i \gamma^i \mathbb{D}_i = {1 \over 2} i \gamma^i \Big[D_i \left((i \gamma^k D_k)^2 -
{1 \over 4} R + M^2 \right)^\alpha + \left((i \gamma^k D_k)^2 - {1 \over 4} R + M^2
\right)^\alpha D_i \Big]
\ee
from which it follows that $i\gamma^i D_i$ and $i \gamma^i \mathbb{D}_i$ do not commute in general.
Their commutator differs from zero by derivatives of the curvature, which provide a geometric
obstruction to find a common system of eigen-spinors $\Psi$. However, the situation simplifies
drastically when the geometry is homogeneous, since $R$ is the same at all points of space and
all its derivatives vanish by definition. Only in this case the two operators exhibit a common
system of eigen-spinors and the corresponding eigen-values are simply related by
\be
Z = \zeta \left(\zeta^2 - {1 \over 4} R + M^2 \right)^\alpha .
\label{rela}
\ee

From now on we restrict attention to homogeneous geometries on $\Sigma_3$ and to be more
specific we consider the class of Bianchi IX metrics on $\Sigma_3 \simeq S^3$ of the following
type
\be
ds^2 = \gamma \Big[ (\sigma^1)^2 + (\sigma^2)^2 + \delta^2 (\sigma^3)^2 \Big] ~,
\label{bianchi}
\ee
using the left--invariant one--forms $\sigma^I$ of the group $SU(2)$ that satisfy the
relations
\be
d\sigma^I + {1 \over 2} {\epsilon^I}_{JK} \sigma^J \wedge \sigma^K = 0 ~.
\ee
Here, we are actually assuming that the geometry has enlarged
isometry $SU(2) \times U(1)$ by taking two of the metric coefficients equal to each
other. Then, the parameters of the metric are the conformal factor $\gamma$ and the
anisotropy parameter $\delta$, which can range from $0$ to $\infty$. These metrics are
often referred in the literature as Berger spheres.

The spectrum of the Dirac operator can be completely determined on such geometries and
the same thing also applies to the spectrum of the more general operators
$i \gamma^i \mathbb{D}_i$. If we were considering the more
general class of Bianchi IX metrics, without imposing axial symmetry,
the eigen-value problem of the Dirac operator would have not been tractable.
For later use we also write the Ricci scalar curvature for this particular
class of three-dimensional metrics \eqn{bianchi},
\be
R= {1 \over 2 \gamma} (4 - \delta^2) ~.
\label{curvae}
\ee

There is a dual basis of vector fields $f_I$ associated to $\sigma^I$ with $<f_I , \sigma^J>
= \delta_I^J$ that satisfy the commutation relations $[f_I , ~ f_J] = - {\epsilon_{IJ}}^K f_K$
and represent the $SU(2)$ Killing vector fields of this particular class of geometries. Both
$\sigma^I$ and $f_I$ can be written in terms of three Euler angles, but the explicit expressions
will not be needed here. Then, the Dirac operator for the class of axially symmetric Bianchi IX
geometries \eqn{bianchi} takes the following form (see also the Appendix)
\be
i \gamma^i D_i = \left(\begin{array}{ccc}
i f_3/\delta &  & i f_1 - f_2 \\
  &  &   \\
i f_1 + f_2 &  & - i f_3/\delta
\end{array} \right) +
{1 \over 4 \delta} (\delta^2 + 2) ~,
\label{dirac}
\ee
where we have taken $\gamma = 1$ without loss of generality. The conformal factor $\gamma$
can be easily reinstated by noting that $i \gamma^i D_i$ scales uniformly as $1/\sqrt{\gamma}$,
but this will not affect the value of the $\eta$-invariant on Berger spheres, since it only
depends on $\delta$. For later use we note that the same thing applies to the operators
$i \gamma^i \mathbb{D}_i$ that scale uniformly by $1/ (\sqrt{\gamma})^{2 \alpha + 1}$ and
their $\eta$-invariant is also be inert to $\gamma$ (actually, in this case, we also have
to rescale appropriately the mass parameter $M$, but it does not really matter since the
$\eta$-invariant will turn out to be independent of $M$ as well).

The spectrum of the Dirac operator \eqn{dirac} on Berger spheres is given by the following
discrete set of eigen-values \cite{hitch} (but see also \cite{bar})
\be
\zeta_{\pm} = {\delta \over 4} \pm {1 \over 2 \delta} \sqrt{4 \delta^2 pq + (p-q)^2} ~;
~~~~~~ (p, q) \in \mathbb{N}^2
\label{sasha1}
\ee
with multiplicities $p+q$ for each pair $(p, q)$. The positive integers $p$ and $q$ are
not ordered, which means that the conjugate pairs $(q, p)$ also yield the same eigen-values
$\zeta_{\pm}$ with the same multiplicities. There are additional eigen--values arising
formally for $q = 0$,
\be
\zeta_0 = {\delta \over 4} + {p \over 2 \delta} ~; ~~~~~~ p \in \mathbb{N} ~,
\label{sasha2}
\ee
but their multiplicity is $2p$ rather than $p$. Thus, the complete spectrum of the Dirac
operator is $\zeta_{\pm}$ and $\zeta_0$ forming a two--dimensional state lattice which is
labeled by the quantum numbers $(p, q)$. The eigen--values $\zeta_+$ and $\zeta_0$ are
positive definite for all $\delta$, whereas $\zeta_-$ can assume positive as well as
negative values by tuning their parameters. Likewise, the spectrum of the more general
operator $i \gamma^i \mathbb{D}_i$ takes the form
\be
Z = \zeta \left(\zeta^2 + {1 \over 8} (\delta^2 - 4) + M^2 \right)^\alpha
\label{loula}
\ee
following from \eqn{rela} and \eqn{curvae} with $\gamma = 1$. The corresponding eigen-values
$Z_{\pm}$ and $Z_0$ are obtained by setting $\zeta = \zeta_{\pm}$ and $\zeta_0$ respectively.

We also note that the sign of the eigenvalues $\zeta$ and $Z$ is the same irrespective
of $\alpha$ and $M$, since
\be
\zeta^2 + {1 \over 8} (\delta^2 - 4) > 0~ , ~~~~~~~~ \forall ~ \delta \in [0, \infty) ~.
\ee
This can be verified directly for $\zeta = \zeta_{\pm}$ or $\zeta_0$ and it is important
for computing the $\eta$-invariant of the corresponding operators by separating the
contribution of their positive and negative modes. Although $\zeta_+$ and $\zeta_0$ are
positive definite for all values of $\delta$, $\zeta_-$ can be either negative or positive
depending upon $\delta$. The eigen-values $\zeta_-$ can also become zero at special values
of $\delta$ for which there exist positive integer solutions $(p, q)$ to the equation
\be
\delta^2 = 2 \sqrt{4pq \delta^2 + (p-q)^2} ~.
\label{harmo}
\ee
The corresponding zero modes $\Psi$ are often referred as harmonic spinors in the
mathematics literature \cite{lichn, hitch}, and, clearly, they can only exist for
$\delta \geq 4$. The presence of zero modes should be properly accounted in the
computation of the $\eta$-invariant.

There are some important points that need to be emphasized before proceeding further.
The anisotropy parameter $\delta$ accounts for the distortion of the three-sphere. The
value $\delta = 1$ corresponds to the round metric on $S^3$. When $\delta$ decreases
the sphere is squashed until it becomes fully flattened in the extreme case $\delta = 0$;
in the latter case $S^3$ collapses to $S^2$ but the geometry is completely regular there.
When $\delta$ increases the sphere is stretched until it becomes singular as
$\delta \rightarrow \infty$. Note that the Ricci scalar curvature \eqn{curvae} is positive
definite for $\delta < 2$ and it turns negative for $\delta > 2$. Then, by Lichnerowicz's
formula \eqn{lichn}, the Dirac operator cannot exhibit any zero modes for
$\delta < 2$. Zero modes are allowed to exist only for $\delta > 2$. Note at this point that
when the curvature of space is positive definite, all eigen-values
$\zeta_-$ are negative and clearly the same thing applies to all eigen-values $Z_-$.
For negative curvature, however, level crossing becomes possible provided that $\delta > 4$.
The critical value $\delta = 4$ is the threshold for the existence of solutions to equation
\eqn{harmo}. This value is far beyond Lichnerowicz's bound $\delta > 2$, and, thus, in this context,
it corresponds to Berger spheres with sufficiently negative curvature. At $\delta = 4$ the
eigen-value $\zeta_-$ with $p=q=1$ becomes zero and then it turns positive for all $\delta > 4$.
As $\delta$ increases further, there are more and more eigen-values $\zeta_-$ that become
positive, until they all undergo level crossing in the extreme limit $\delta \rightarrow \infty$.

The computation of the $\eta$-invariant $\eta_{\mathbb{D}} (S^3)$ relies on the separation of
positive and negative modes. First, we assume that $\delta < 4$ so that the eigen-values
$Z_-$ are all negative, whereas $Z_+$ and $Z_0$ are all positive. Later we will extend the
result to Berger spheres with $\delta > 4$ by counting the number of modes that have
undergone level crossing. Thus, according to definition,
\be
\eta_{\mathbb{D}} = \lim_{s \rightarrow 0}
\eta_{\mathbb{D}} (s) = \lim_{s \rightarrow 0} \sum_{{\rm eigenvalues}} ({\rm sign} Z)
|Z|^{-s} ,
\ee
we have to evaluate the following sums when $\delta < 4$,
\ba
\eta_{\mathbb{D}} (s) & = & \sum_{p, q >0} (p+q) \left(\left({\delta^2 \over 2} + X
\right) \Big[\left({\delta^2 \over 2} + X\right)^2 + {\delta^2 \over 2} (\delta^2 - 4)
+ 4 \delta^2 M^2 \Big]^\alpha \right)^{-s} - \nonumber\\
& & \sum_{p, q >0} (p+q) \left(\left(-{\delta^2 \over 2} + X \right)
\Big[\left(-{\delta^2 \over 2} + X \right)^2 + {\delta^2 \over 2} (\delta^2 - 4)
+ 4 \delta^2 M^2 \Big]^\alpha \right)^{-s} + \nonumber\\
& & \sum_{p>0} 2p \left(\left({\delta^2 \over 2} + p \right)
\Big[\left({\delta^2 \over 2} + p \right)^2 + {\delta^2 \over 2} (\delta^2 - 4)
+ 4 \delta^2 M^2 \Big]^\alpha \right)^{-s} ,
\label{main}
\ea
setting $X = \sqrt{4 \delta^2 pq + (p-q)^2}$ for notational convenience.
The first line refers to the contribution of $Z_+$, the second to $Z_-$ and the third
to $Z_0$. Note that in writing down \eqn{main} all eigen-values $\zeta$ in \eqn{loula}
have been rescaled by a factor of $2 \delta$, which is irrelevant, however, for the
final result that is obtained at $s=0$.

Next, we compute the individual sums by expanding all fractions in powers of
$\delta$ (up to the appropriate order) and then set $s=0$.
The contribution of the modes $Z_0(\delta)$ is the easiest to evaluate. Using the
power series expansion
\ba
& & \left(\left({\delta^2 \over 2} + p \right)
\Big[\left({\delta^2 \over 2} + p \right)^2 + {\delta^2 \over 2} (\delta^2 - 4)
+ 4 \delta^2 M^2 \Big]^\alpha
\right)^{-s} = \nonumber\\
& & {1 \over p^{(2 \alpha + 1) s}} \Big[1 - s \delta^2 \left({2 \alpha + 1 \over 2p}
+ {2 \alpha (2M^2 -1) \over p^2} \right)
+ {s\delta^4 \over 8p^2} \left((s+1) (2 \alpha + 1)^2 - 2 \alpha (2 \alpha + 3) \right)
\nonumber\\
& & + \cdots \Big]
\label{calc1}
\ea
we find that the last group of terms in \eqn{main} assumes the following expansion
written in terms of Riemann zeta--functions,
\ba
I_0 (s) & = & 2 \zeta ((2 \alpha + 1) s-1) - s \delta^2 \Big[(2 \alpha +1) \zeta ((2 \alpha + 1)s)
+ 4 \alpha (2M^2 -1) \zeta ((2 \alpha +1)s + 1) \Big] \nonumber\\
& & + {s \delta^4 \over 4} \Big[(s+1)(2 \alpha + 1)^2 - 2 \alpha (2\alpha + 3) \Big]
\zeta ((2 \alpha + 1)s+1) + \cdots ~.
\ea
The terms that are omitted vanish at $s=0$ as they contain the factor $s\zeta ((2 \alpha + 1) s+n)$
with integer $n \geq 2$ and $\zeta (s)$ is absolutely convergent for ${\rm Re} ~ s > 1$;
they include a term of the form $\delta^4 /p^3$ in \eqn{calc1}
as well as all terms of order $\delta^6$ and higher which come multiplied with
$1/p^{n+1}$ with $n \geq 2$. Then, taking into account that $\zeta (-1) = -1/12$,
$\zeta(0) = -1/2$ and that $\zeta (s)$ has a simple pole at $s=1$ with residue $1$ (and,
thus, $(2 \alpha + 1) s \zeta ((2 \alpha + 1) s +1)$ equals $1$ at $s=0$), we obtain
\be
I_0 (0) = -{1 \over 6} - {4 \alpha \over 2 \alpha + 1} (2M^2 - 1) \delta^2 -
{2 \alpha - 1 \over 4 (2 \alpha + 1)} \delta^4 ~.
\ee

The contribution of the modes $Z_{\pm} (\delta)$ to the $\eta$--invariant is more difficult
to extract because of the double sums that are involved. Again, we expand the fractions in
power series of $\delta$
\ba
& & \left(\left(\pm {\delta^2 \over 2} + X \right) \Big[\left(\pm {\delta^2 \over 2} +
X \right)^2 + {\delta^2 \over 2} (\delta^2 - 4) + 4 \delta^2 M^2 \Big]^\alpha \right)^{-s}
= \nonumber\\
& & {1 \over X^{(2 \alpha + 1)s}} \Big[1 - s \delta^2 \left(\pm {2 \alpha +1 \over 2X} +
{2 \alpha (2M^2 - 1) \over X^2} \right) + \nonumber\\
& & s \delta^4 \left({(s+1)(2\alpha + 1)^2 - 2\alpha (2\alpha + 3) \over 8X^2} \pm
{\alpha [(s+1)(2\alpha + 1) - (2\alpha -1)](2M^2 -1)) \over X^3}
\right) \mp \nonumber\\
& & {s \delta^6 \over 48 X^3} \left(2\alpha (2\alpha -1) (2\alpha + 7) +
(s+1)(2 \alpha +1) [(s+2) (2\alpha +1)^2 - 6 \alpha (2 \alpha + 3)] \right) \nonumber\\
& & + \cdots \Big]
\label{calc2}
\ea
omitting all terms of the form $1/X^{n+2}$ with $n \geq 2$ that do not contribute to the
final result when $s=0$, as will be seen shortly. Then, the first two sums in the general
expression \eqn{main} for $\eta_{\mathbb{D}}(s)$, which we denote respectively by
$I_{\pm} (s)$, are combined together as follows,
\ba
& & I_+ (s) - I_- (s) = -(2 \alpha +1)s \delta^2 f\left({(2 \alpha +1)s+1 \over 2}\right)
- \nonumber\\
& & 2s f\left({(2 \alpha +1)s+3 \over 2}\right)
\Big\{\alpha [2 \alpha - 1 - (s+1)(2 \alpha + 1)] (2M^2 -1) \delta^4 + \nonumber\\
& & {\delta^6 \over 48} \left(2 \alpha (2 \alpha -1) (2 \alpha + 7) + (s+1)(2 \alpha +1)
[(s+2) (2 \alpha +1)^2 - 6 \alpha (2 \alpha + 3)] \right) \Big\} \nonumber\\
& & + \cdots ~,
\ea
setting for convenience
\be
f(s) = \sum_{p, q >0} {p+q \over X^{2s}} = \sum_{p, q >0} {p+q \over [4 \delta^2 pq +
(p-q)^2]^s} ~.
\ee

The function $f(s)$ is absolutely convergent on the complex $s$--plane for ${\rm Re} ~ s > 3/2$,
which justifies the suppression of the higher order terms in the power series expansion above
(all such terms are multiplied with $s$ and vanish at $s=0$). Furthermore, the residues of the
function $f(s)$ at the two special points $s= 1/2$ and $3/2$, which are relevant for the
calculation, are (see, for instance, \cite{habel})
\be
{\rm res} f(s) |_{s=1/2} = {\delta^2 - 1 \over 6} ~, ~~~~~~
{\rm res} f(s) |_{s=3/2} = {1 \over 2 \delta^2} ~,
\ee
and, therefore, the resulting contribution of these terms to the $\eta$-invariant is
\be
I_+ (0) - I_- (0) = {\delta^2 \over 3} \left(1 + {12 \alpha \over 2 \alpha + 1} (2M^2 -1) \right)
- {\delta^4 \over 3} \left(1 - {10 \alpha -1 \over 4(2 \alpha + 1)} \right) .
\ee

Putting all together, it turns out that the $\eta$-invariant is independent of the exponent
$\alpha$ and the scale parameter $M$. The final result reads
\be
\eta_{\mathbb{D}} (0) = I_+ - I_- + I_0 = - {1 \over 6} (1 - \delta^2)^2 .
\ee
As such, it coincides with the $\eta$-invariant of the Dirac operator on Berger spheres with
anisotropy parameter $\delta$ and it is related to the gravitational Chern-Simons action as
follows (see also the Appendix),
\be
{1 \over 192 \pi^2} W_{\rm CS} ({\rm Berger}) = {1 \over 12} (2 - 2\delta^2 + \delta^4) =
- {1 \over 2} \eta_{\mathbb{D}} (0) + {1 \over 12} ~,
\ee
in agreement with formula \eqn{moulia} for the Dirac operator. This proves our claim \eqn{eetaa}
for the $\eta$-invariant on Berger spheres with $\delta < 4$ and suffices to show that the
coefficient of the anomaly is indeed universal.

Next, we examine the case $\delta > 4$ that involves level crossing. Every time an eigen-value
$Z_-$ crosses from negative to positive values, $\eta_{\mathbb{D}}$ jumps by $2$ to
$\eta_{\mathbb{D}} + 2$, since the spectral asymmetry, as calculated for $\delta < 4$, changes
by $2$. Let us denote by
\be
{\cal C}(\delta_{\rm c}) = \{(p, q) \in \mathbb{N}^2 ; ~~ \delta_{\rm c}^2 = 2 \sqrt{4pq
\delta_{\rm c}^2 + (p-q)^2} \}
\ee
the set of positive integers $(p, q)$ that account for harmonic spinors at each one of the
special values $\delta_{\rm c} < \delta$ arising for fixed $\delta > 4$. Since the zero
modes at $\delta_{\rm c}$ have multiplicity $p+q$, the quantity
\be
S(\delta) = \sum_{\delta_{\rm c} < \delta} \Big[\sum_{(p, q) \in {\cal C}(\delta_{\rm c})}
(p+q) \Big]
\ee
provides the total number of modes that crossed from negative to positive values as $\delta$
was varying from $\delta < 4$ to any given value $\delta > 4$.
Thus, for $\delta > 4$, the $\eta$--invariant of the operator $\mathbb{D}$ on Berger spheres is
shifted by twice the number of these modes and equals
\be
\eta_{\mathbb{D}} = - {1 \over 6} (\delta^2 - 1)^2 + 2 S(\delta) ~.
\label{geneta}
\ee
$S(\delta)$ is also inert to the parameters $\alpha$ and $M$ which shows that level crossing
is the same as for the Dirac operator. It can be easily seen that $S(\delta)$ is always an
even integer, but there is no closed expression for it for general values of $\delta$.

The results we have presented here generalize our earlier work on the subject \cite{dieter}
to all values of the anisotropy scaling parameter $z$ and show that the axial anomaly
is an infra-red phenomenon, as in relativistic quantum field theory. The universality of the
$\eta$-invariant also shows that the parity breaking piece of the induced fermion action
in three-dimensions is provided by the Chern-Simons term with the same coefficient as in
\eqn{mitia} irrespective of the order of the fermion operator $i \gamma^i \mathbb{D}_i$.
Finally, the model geometries that were used to perform the calculation provide
simple examples for which the index of the four-dimensional fermion operator can be
computed explicitly by spectral flow as ${\rm Ind} ({\cal D}) = \pm S(\delta)$ (the sign
depends on the choice of orientation). In the next section, we consider four-dimensional
backgrounds that can give rise to violations of the chiral charge conservation law
in gravitational theories, and, thus, lead to baryon and lepton number
violation when the index differs from zero (note at this end that the axial anomaly
will not obstruct the conservation law $\Delta Q_5 = 0$ if the index is zero).

\section{Applications to non-relativistic theories of gravity}
\setcounter{equation}{0}

So far the geometry has been treated as background for fermion propagation without
reference to any metric field equations. Space-times of the form $I \times \Sigma_3$
with Euclidean line element \eqn{spatim} will lead to violations of the chiral charge
conservation if the gravitational field equations for the metric $g_{ij} (t, x)$ induce
level crossing for the fermion operator on $\Sigma_3$ as $t$ varies in
$I$. In Einstein gravity this cannot happen for various reasons
(see, for instance, \cite{gibbons}), but in non-relativistic gravitational theories
of Ho\v{r}ava-Lifshitz type \cite{horava} there are regular solutions that can support
such exotic effects.

For definiteness, we concentrate on gravitational instanton solutions in $3+1$ dimensions
with $SU(2) \times U(1)$ isometry for which $\Sigma_3 \simeq S^3$ is a Berger sphere
with metric \eqn{bianchi}. In Einstein gravity these are the familiar Taub-NUT and
Eguchi-Hanson instantons for which $I$ is the semi-infinite line and the anisotropy
parameter $\delta$ changes between $0$ and $1$ as $t$ varies between the two end-points
of $I$; in these cases one side of $I$ terminates at a removable singularity, which
is a nut or a bolt respectively. The field equations do not allow $\delta$ to exceed
$1$ without ruining the regularity and completeness of the four-dimensional metrics.
Thus, in this case, level crossing cannot occur and the index of the four-dimensional
Dirac operator on such metric backgrounds is zero. More generally, for non-compact
four-metrics with non-negative Ricci scalar curvature, there can be no bound state
solutions of the Dirac equation; if such states existed, they would have been covariantly
constant, and, hence, non-normalizable \cite{lichn} leading to contradiction.

On the other hand, in Ho\v{r}ava-Lifshitz theory with detailed balance (meaning that
the potential term of the action is derivable from a local superpotential functional
$W [g]$) the instantons are eternal solution of the gradient flow equation for the
three-metric $g$ derived from $W$ \cite{BBLP}. In this case, $I \simeq R$ and the
instanton interpolates between two distinct vacua of the three-dimensional action $W$
(viewed as fixed points of the gradient flow equation) as $t$ ranges from $-\infty$
to $+\infty$. Typically, on $S^3$, $W[g]$ admits a maximally symmetric vacuum for
which $\delta =1$, but there should be other less symmetric vacua that coexist and provide
the other end-point of the instantons. Then, within our Bianchi IX mini super-space
model, we pose the following problem: under what conditions $W[g]$ admits
another less symmetric vacuum with $\delta > 4$, in order to have violation of chiral
charge conservation induced by instantons, and what is the phase of the non-relativistic
gravitational theory that admits such instantons as bona fide Euclidean solutions?
The results we describe below show that $Q_5$ cannot be always conserved in
Ho\v{r}ava-Lifshitz gravity with anisotropy scaling exponent $z \geq 3$. For lower
values of $z$ there seems to be no window in the space of couplings for having
non-conservation of chiral charge. Curiously, Ho\v{r}ava-Lifshitz gravity becomes
power counting renormalizable only for $z \geq 3$, which was the reason to view it as
substitute of general relativity in the deep ultra-violet regime in the first place
\cite{horava}. Thus, it is interesting to study the problem of chiral symmetry
breaking in these models.

Recall that the superpotential functional $W[g]$ is taken to be the action of
three-dimensional gravity on $\Sigma_3$ augmented with higher derivative terms. A simple
choice is provided by topologically massive gravity \cite{jackiw} containing the
gravitational Chern-Simons term \eqn{csact} and a three-dimensional cosmological
constant $\Lambda_{\rm w}$,
\be
W_{\rm TMG} [g] = {2 \over \kappa_{\rm w}^2} \int_{\Sigma_3} d^3x \sqrt{{\rm detg}} ~
(R-2\Lambda_{\rm w}) + {1 \over 2 \omega} W_{\rm CS} [g] ~.
\ee
The corresponding theory of Ho\v{r}ava-Lifshitz gravity has anisotropy scaling parameter
$z=3$ unless the Chern-Simons term is absent in which case $z=2$. The vacua of topologically
massive gravity satisfy the following field equations, which are split on purpose into
traceless and trace parts,
\be
R_{ij} - {1 \over 3} R g_{ij} - {\kappa_{\rm w}^2 \over \omega} C_{ij} = 0 ~~~~
{\rm and} ~~~~ R = 6 \Lambda_{\rm w} ~,
\ee
where $C_{ij}$ is the Cotton tensor of the three-metric $g$.

For the class of Bianchi IX metrics \eqn{bianchi}, the traceless part of the field
equations reduce to a single algebraic relation among the anisotropy parameter
$\delta$ and the conformal factor $\gamma$,
\be
{1 - \delta^2 \over 3} + \kappa_{\rm w}^2 {\delta (1 - \delta^2) \over 2 \omega
\sqrt{\gamma}} = 0 ~.
\ee
Thus, apart from the maximally symmetric solution $\delta = 1$, there is one more
solution for
\be
{1 \over 3} + {\kappa_{\rm w}^2 \delta \over 2 \omega \sqrt{\gamma}} = 0
\label{mailei}
\ee
provided that the Chern-Simons coupling $\omega$ is negative (for a given choice of
orientation on $S^3$). The two metrics can coexist as vacua of topologically massive
gravity only when $\Lambda_{\rm w}$ is non-negative. This follows from the trace
part of the field equations $R = 6 \Lambda_{\rm w}$ using \eqn{curvae}. Then, the
corresponding instanton of Ho\v{r}ava-Lifshitz gravity interpolates between two
Berger spheres with the same non-negative curvature, and, thus, there can be no
net level crossing in the spectrum of the three-dimensional fermion operator. The
only way to have level crossing is to be in a unimodular phase of the theory in
which the trace part of the field equations decouples from the dynamics and the two
end-points are allowed to have different curvature (positive for $\delta = 1$
and sufficiently negative for the other solution \eqn{mailei}).

This possibility is naturally realized when the generalized DeWitt metric in super-space
arising in the canonical formulation of the theory \cite{horava}
\be
{\cal G}^{ijkl} = {1 \over 2} (g^{ik} g^{jl} + g^{il} g^{jk}) - \lambda g^{ij} g^{kl}
\ee
has $\lambda = 1/3$ and projects any tensor to its traceless part
(${\cal G}_{ijkl}$ has similar form and properties). Then, the volume of space is
preserved by the dynamics in $3+1$ dimensions. Since the volume of the Berger sphere
\eqn{bianchi} is fixed ${\rm Vol} (S^3) = 16 \pi^2 \delta \gamma^{3/2}$,
we can rewrite the defining equation of the anisotropic fixed point \eqn{mailei}
as follows by eliminating $\gamma$,
\be
{\rm Vol} (S^3) = 54 \pi^2 \delta^4 \left(-{\kappa_{\rm w}^2 \over \omega}\right)^3 .
\ee
Thus, in this case, the index of the Dirac-Lifshitz operator will be non-zero on the
corresponding gravitational instanton background provided that $\delta > 4$, which
in turn imposes a lower bound on the volume of space written in terms of the ratio
of the couplings $\kappa_{\rm w}^2$ to $\omega$ as follows,
\be
{\rm Vol} (S^3) > 13824 \pi^2 \left(-{\kappa_{\rm w}^2 \over \omega}\right)^3 ,
\ee
reproducing the results reported in \cite{dieter}. Note that the second fixed point
\eqn{mailei} is absent when there is no Chern-Simons term in $W[g]$, and, hence, there
is no instanton to induce violation of chiral charge conservation when the anisotropy
scaling exponent is lowered to $z=2$.

Next, we consider the case of $z=4$ Ho\v{r}ava-Lifshitz gravity in $3+1$ dimensions
(see, for instance, \cite{cai}) choosing as $W[g]$ the action of new massive gravity on
$\Sigma_3$, \cite{holm},
\be
W_{\rm NMG} [g] = {2 \over \kappa_{\rm w}^2} \int_{\Sigma_3} d^3x \sqrt{{\rm detg}} ~
(R-2\Lambda_{\rm w}) + {1 \over m^2} \int_{\Sigma_3} d^3x \sqrt{{\rm detg}}
\left(R_{ij} R^{ij} - {3 \over 8} R^2 \right) .
\label{nmgr}
\ee
We want to examine if the results described above persist in this case, leaving again a
window for violation of chiral charge conservation by gravitational instanton effects.
The choice \eqn{nmgr} is not the most general one that can be made. One may augment it
with a gravitational Chern-Simons term, as will be discussed briefly later, and also
change the relative coefficient of the quadratic curvature terms in \eqn{nmgr}, which,
however, will not be discussed here for simplicity (such a generalization spoils some
nice properties of new massive gravity as higher curvature extension of three dimensional
Einstein gravity, but it is legitimate in the context of $z=4$ Ho\v{r}ava-Lifshitz gravity).

For Berger spheres, the traceless part of the classical equations of motion of new
massive gravity reduce consistently to the following algebraic relation (see \cite{sourdis}
for details),
\be
{1 - \delta^2 \over 3} + \kappa_{\rm w}^2 {(1 - \delta^2) (4 - 21 \delta^2) \over 48 m^2
\gamma} = 0 ~.
\ee
Thus, apart from the maximally symmetric solution $\delta = 1$ we also have another solution
for
\be
1 + {\kappa_{\rm w}^2 (4 - 21 \delta^2) \over 16 m^2 \gamma} = 0 ~.
\label{rauta}
\ee
Since we are ultimately interested in the coexistence of vacua with $\delta > 4$, we impose
the restriction $m^2 > 0$ on the coupling of the quadratic curvature terms (otherwise, if
$\delta^2 < 4/21$, the existence of a second solution will require $m^2 < 0$). The trace
part of the field equations yields
\be
{4-\delta^2 \over 2 \gamma} = 6 \Lambda_{\rm w} + {\kappa_{\rm w}^2 \over 64m^2 \gamma^2}
(3\delta^2 - 4)(7\delta^2 -4) ~.
\label{trimae}
\ee
Setting $\delta = 1$, we find that $\Lambda_{\rm w}$ is non-negative for $m^2 >0$. The
other solution \eqn{rauta} does not satisfy the trace relation for $\delta >4$, since
the left-hand side will be negative and the right-hand side positive. Thus, the only way
that both solutions can coexist when $\delta > 4$ is to decouple the trace part of the
field equations from the dynamics.

As before, this only happens in the unimodular phase of the theory corresponding to the
choice $\lambda = 1/3$ in the DeWitt metric. Then, the volume of space remains fixed and
can be expressed in terms of $\delta$ by eliminating $\gamma$ from the fixed point \eqn{rauta}.
We have, in particular, the following relation
\be
{\rm Vol} (S^3) = {\pi^2 \kappa_{\rm w}^3 \over 4 m^3} \delta  (21 \delta^2 -4)^{3/2}
> 8 \pi^2 (83)^{3/2} {\kappa_{\rm w}^3 \over m^3} ~,
\ee
using $\delta > 4$ to establish the lower bound in the volume of space in terms of the
couplings. This inequality provides the window for having violation of chiral charge
conservation in the corresponding $3+1$ theory of Ho\v{r}ava-Lifshitz gravity.

Further generalizations arise by adding the gravitational Chern-Simons term to
$W_{\rm NMG} [g]$ that yield the so called three-dimensional generalized new massive
gravity \cite{holm}. This theory has a maximally symmetric solution with $\delta =1$.
Other Berger sphere solutions satisfy the traceless part of the field equations,
which boil down to
\be
{1 \over 3} + {\kappa_{\rm w}^2 \delta \over 2 \omega \sqrt{\gamma}} +
{\kappa_{\rm w}^2 (4 - 21 \delta^2) \over 48 m^2 \gamma} = 0 ~.
\label{rauta2}
\ee
For simplicity, we assume $m^2 > 0$ so that there is only one admissible solution of
equation \eqn{rauta2}, whereas $\omega$ can take either positive or negative values.
The trace part of the field equations is the same as before, \eqn{trimae}, because the
Cotton tensor is traceless. Thus, for $m^2 > 0$, \eqn{trimae} cannot be possibly
satisfied for both $\delta = 1$ and $\delta > 4$ and we are forced again to consider
the unimodular variant of the theory that corresponds to $\lambda = 1/3$.
Then, the volume of space is preserved throughout the evolution and it is given by
\ba
{\rm Vol}(S^3) & = & {1 \over 4} \pi^2 \delta^4 \left(-{3 \kappa_{\rm w}^2 \over \omega}
+ \sqrt{\left({3 \kappa_{\rm w}^2 \over \omega}\right)^2 + {\kappa_{\rm w}^2 \over m^2}
\left(21 - {4 \over \delta^2} \right)} ~ \right)^3 \nonumber\\
& > & 64 \pi^2 \left(-{3 \kappa_{\rm w}^2 \over \omega}
+ \sqrt{\left({3 \kappa_{\rm w}^2 \over \omega}\right)^2 +
{83 \kappa_{\rm w}^2 \over 4m^2}} ~ \right)^3 .
\ea
The lower bound is obtained by demanding $\delta > 4$ so that chiral charge conservation
can be violated in the corresponding Ho\v{r}ava-Lifshitz gravity and it encompasses the
two simpler cases that were discussed before. The analysis is more intricate when
$m^2 < 0$, as there can be more than one vacua with $\delta \neq 1$, and, hence, more
instantons that interpolate between them; the details are left to the interested reader.

Similar considerations apply to theories based on Born-Infeld generalizations of
three-dimensional gravity \cite{tekin} that include higher and higher derivative terms.
The anisotropy scaling exponent $z$ of the corresponding Ho\v{r}ava-Lifshitz  theory
increases accordingly and there can be several Berger sphere solutions that coexist
and support the end-points of instantons. We expect that the general features of our
results will remain qualitatively the same in these cases too, leaving windows in the
space of gravitational coupling parameters for having chiral symmetry breaking by
gravitational instanton effects. This novelty differentiates the non-relativistic
gravitational theories of Lifshitz type from Einstein gravity and will have
phenomenological consequences in bouncing models of the early universe if
Ho\v{r}ava-Lifshitz gravity plays a role. We also expect these results to be valid more
generally beyond the mini-superspace model geometries, which were only used here for
illustrative purposes.

In all examples that were considered above, it was found that non-conservation of
chiral charge can only occur in the phase of Ho\v{r}ava-Lifshitz gravity with
$\lambda = 1/3$, where the volume of space decouples from the dynamics. This phase
of the theory is not inflicted with many of the pathologies that otherwise haunt
non-relativistic models of gravitation. It may very well be that only this phase
becomes relevant in the deep ultra-violet regime, as substitute for ordinary gravity,
and not its variants with $\lambda \neq 1/3$. Flowing from the ultra-violet to the
infra-red is a difficult problem that could be properly investigated within the
asymptotic safety scenario for quantum gravity (for a review see, for instance,
\cite{martin} and references therein) treating $\lambda$ as a coupling that undergoes
renormalization. Proper treatment of the problem may also require turning on other
operators that are lying outside the class of Ho\v{r}ava-Lifshitz models to ensure
well-behaved transition to ordinary large scale physics. All these issues are currently
under consideration.

\section{Conclusions}
\setcounter{equation}{0}

We have investigated the occurrence of axial anomalies in non-relativistic fermion theories
of Lifshitz type which are minimally coupled to gauge and/or gravitational fields in
$3+1$ space-time dimensions. Our main conclusion is that the results are identical to the
relativistic case, hereby generalizing earlier work on subject \cite{wadia, dieter} to
theories with arbitrary anisotropy scaling exponents. This is not surprising in retrospect
conforming with the general idea that the axial anomaly (unlike others) is an infra-red
phenomenon. Also, the universal form of the $\eta$-invariant can be intuitively understood
(prior to zeta-function regularization) by noting that the operator $(-D^2 + M^2)^{\alpha}$ is
positive definite; hence, the number of positive and negative modes is the same for both
operators $i \gamma^i D_i$ and $i \gamma^i \mathbb{D}_i$. It was
confirmed that the same conclusion holds for the regulated $\eta$-invariants.

Although the result has its own value, we have also considered some
applications in the context of non-relativistic gravitational theories of Lifshitz type
and found that (unlike ordinary gravity) violation of chiral charge conservation can
be induced by instantons - and possibly other configurations - in certain regions of
their parameter space.

There are a few technical questions that remain open and should be addressed carefully
elsewhere. It is not yet clear whether there are any restrictions on the allowed
values of the anisotropy scaling parameter $z$ that insure self-adjointness
of the Dirac-Lifshitz operator in the strict mathematical sense. Also, it is interesting
to generalize our results to higher dimensions and compare with the anomalies of the
corresponding relativistic quantum field theories. Finally, it remains to
explore the role of anomalies in condensed matter physics systems that exhibit quantum
criticality.

\newpage
%\vskip1cm

\centerline{\bf Acknowledgements}
\noindent
This work was initiated during the 41st Summer Institute at Ecole Normale Sup\'erieure
in Paris in August 2011, carried at EISA Summer Institute at Corfu and completed during
a visit at CERN. I thank \'Edouard Brezin for posing the general question and his
encouragement to perform the calculation. I also thank the organizers of the ENS summer
institute for their kind invitation and financial support as well as the other participants
for useful discussions on the subject of this paper. Special thanks are due to
Luis Alvarez-Gaum\'e, Costas Kounnas, Dieter L\"ust and Arkady Tseytlin for
related scientific discussions and comments. The hospitality and financial support of the
theory group at CERN is also gratefully acknowledged.

\appendix
\section{Chern-Simons action for Berger spheres}
\setcounter{equation}{0}

In this appendix we provide the components of the connection one-form of the Berger sphere
with metric $ds^2 = \gamma \left((\sigma^1)^2 + (\sigma^2)^2 + \delta^2 (\sigma^3)^2 \right)$.
These expressions are needed for writing down the Dirac operator on such backgrounds as well as
for evaluating the gravitational Chern-Simons action $W_{\rm CS}({\rm Berger})$ used in the
main text. We have, in particular,
\be
{\omega^1}_2 = {\delta^2 - 2 \over 2} \sigma^3 ~, ~~~~~~ {\omega^1}_3 = {\delta \over 2}
\sigma^2 ~, ~~~~~~ {\omega^2}_3 = -{\delta \over 2} \sigma^1 ,
\ee
noting that they are all independent of the scale factor $\gamma$.

The gravitational Chern-Simons action can be easily evaluated in steps as follows,
\ba
W_{\rm CS} & = & \int_{S^3} {\rm Tr} \left(\omega \wedge d \omega + {2 \over 3}
\omega \wedge \omega \wedge \omega \right) \nonumber\\
& = & 2 \int_{S^3} \left({\omega^1}_2 \wedge d {\omega^2}_1 +
{\omega^1}_3 \wedge d {\omega^3}_1 + {\omega^2}_3 \wedge d {\omega^3}_2 +
2 {\omega^1}_2 \wedge {\omega^2}_3 \wedge {\omega^3}_1 \right) \nonumber\\
& = & (\delta^4 - 2 \delta^2 + 2) \int_{S^3} \sigma^1 \wedge \sigma^2 \wedge \sigma^3 ,
\ea
using along the way $d \sigma^1 + \sigma^2 \wedge \sigma^3 = 0$ and cyclic permutations.
Since the integral over $S^3$ of the volume form $\sigma^1 \wedge \sigma^2 \wedge \sigma^3$
equals $16 \pi^2$, we obtain the final result
\be
W_{\rm CS}({\rm Berger}) = 16 \pi^2 (\delta^4 - 2 \delta^2 + 2) ~.
\ee

The extrema of this action, as function of $\delta \in [0, \infty)$, occur at $\delta = 1$ 
and $\delta = 0$ in accordance with the fact that the round sphere and the fully 
squashed Berger sphere are the only metrics in this class that are conformally flat. 
The minimum corresponds to $\delta =1$, whereas $\delta =0$ is a local maximum. 

%\newpage

\end{document}